\documentclass[twocolumn]{autart}    
\usepackage{natbib}
\usepackage{har2nat}
\usepackage{amsmath}
\usepackage{amssymb}
\usepackage{bbm}
\usepackage{color}
\usepackage{epstopdf}
\usepackage{siunitx}
\usepackage{bm}
\usepackage{graphicx}
\graphicspath{{figures/}}

\def\twon #1{\|#1\|}
\def\rainfty{\rightarrow\infty}

\def\sgn{\text{sgn}}
\def\sat{\text{sat}}

\def\diag{\text{diag}}

\def \qed {\hfill \vrule height6pt width 6pt depth 0pt}
\def\bee{\begin{equation}}
\def\ene{\end{equation}}
\def\beq{\begin{eqnarray}}
\def\enq{\end{eqnarray}}
\def\bmatri{\begin{bmatrix}}
\def\ematri{\end{bmatrix}}

\begin{document}
	\begin{frontmatter} 
		\title{Target Encirclement  with any Smooth Pattern Using Range-only Measurements \thanksref{footnoteinfo}} \thanks[footnoteinfo] {This work was supported  in part by the National Natural Science Foundation of China under Grant  61722308 and in part by the National Key  Research and Development Program of China under Grant 2016YFC0300801.} 
		\author{Fei~Dong},
		\ead{dongf17@mails.tsinghua.edu.cn}
		\author{Keyou~You\corauthref{cor}},
		\corauth[cor]{Corresponding author} 
		\ead{youky@tsinghua.edu.cn}
		\author{Shiji~Song}
		\ead{shijis@tsinghua.edu.cn}
		\address{Department of Automation and BNRist, Tsinghua University, Beijing, 100084, China.}
		
		\begin{abstract} 
This paper proposes a coordinate-free controller to drive a mobile robot to encircle a target at unknown position by only using range measurements. Different from the existing works,   a backstepping based controller is proposed to encircle the target with zero steady-state error for any desired smooth pattern. Moreover, we show its asymptotic exponential convergence under a fixed set of control parameters, which are independent of the initial distance to the target.  The effectiveness and advantages of the proposed controller are validated via simulations.   

\end{abstract}
\end{frontmatter}
\section{Introduction}

Target tracking with robots has been widely applied in both military and civilian fields, such as border patrol, convey protection, and aerial surveillance, and has attracted considerable research attention in decades.
One of the tracking patterns is {\em target encirclement} \citep{Marasco2012Model}, which requires that a tracking robot encloses the target with flexible commands to neutralize the target by restricting its movement. Particularly, {\em circumnavigation} refers to that the robot exactly slides on a circle centered at the target \citep{Shames2012Circumnavigation,Matveev2011Range}. Many works have focused on this problem, see e.g., \citet{Frew2007Lyapunov,Deghat2014Localization,Hafez2014Using,Cao2015UAV,Zhang2017Unmanned,Xiao2017Target,Yu2017Target} and references therein. However, their controllers cannot be applied to the target encirclement of this work.

If the states (position, velocity, course, etc.) of both the robot and target are available, a Lyapunov guidance vector fields method is proposed by \citet{Lawrence2003Lyapunov} and adopted in \citet{Frew2008Coordinated}. Moreover, both the backstepping control method and the sliding mode control method have been introduced in \citet{ZHONG1997Tracking} and \citet{Lee2013Tracking}, respectively. Clearly, they are coordinate-based and cannot be applied in the GPS-denied environment. 

If the robot state is available but the target state is unknown, e.g., the target is an intruder, the target position can be estimated based on sensor measurements, such as range-only \citep{Shames2012Circumnavigation}, bearing-only \citep{Deghat2014Localization,Zheng2015Enclosing}, or received signal strength \citep{Hu2011Energy}. For a stationary target, an adaptive localization algorithm is devised using range-only measurements in \citet{Shames2012Circumnavigation}.
By using a single vision camera, a vision-based motion estimator and an extended Kalman filter are designed in \citet{Dobrokhodov2008Vision} and \citet{Zhang2010Vision}, respectively. Note that it is impossible to locate the target if the robot state is unknown.   

If neither the robot state nor the target position is available, e.g., the robot is an underwater or indoor vehicle, this problem becomes much more difficult. A sliding mode approach is proposed in \citet{Matveev2011Range} to solve the circumnavigation problem. Although the chattering phenomenon can be eliminated or reduced, their approach cannot achieve zero steady-state. Besides, it requires a particular assumption that the initial position of the robot is sufficiently far away from the target.
A range-only controller is devised in \citet{Milutinovic2014Coordinate,Milutinovi2017Coordinate}, whose control parameters strongly depend on the initial distance to the target. In addition, \citet{Cao2015UAV} introduces a geometrical guidance law, whose idea is to drive the robot towards the tangent point of an auxiliary circle. However, this method involves the computation of trigonometric and inverse trigonometric functions. Since there is no control input to the robot if it enters this auxiliary circle, this may result in large overshoots. To solve it, a switching idea is adopted in \citet{Zhang2017Unmanned} where the switching is performed based on the tracking error. 

Note that all the aforementioned range-only based controllers are only applicable to the circumnavigation problem of a stationary target. Clearly, this pattern cannot adapt to the time-varying/complicated environment, e.g. obstacle avoidance, circumnavigation of multiple targets, moving target, and etc.  Thus, we are interested in the so-called target encirclement with smooth time-varying reference commands in this work.  To this purpose, we propose a coordinate-free controller with range-only measurements by exploiting the backstepping control. In a preliminary version of this work \citep{dong2019circumnavigation}, we have shown its effectiveness for the circumnavigation problem. For a time-varying reference command, our controller naturally contains its time derivative  in the recursively backward process. Thus, our approach can guarantee global convergence and exponential stability with zero steady-state error. Moreover, the control parameters are independent of the initial state due to the use of a saturation function in the backstepping control. The effectiveness and advantages of the proposed controller are validated via simulations.

The rest of this paper is organized as follows. In Section \ref{sec2}, the problem under consideration is formulated in details.  In Section \ref{sec3}, the proposed controller is given by exploiting the backstepping control method. In Section \ref{sec4}, we prove the stability and convergence of the proposed controller for the circumnavigation problem. Target encirclement with smooth time-varying commands is shown in Section \ref{sec7}. Simulations are included in Section \ref{sec5}, and some concluding remarks are drawn in Section \ref{sec6}.

\section{Problem Formulation} \label{sec2}

Let $\bm p_o:=[x_o,y_o]'$ be the {\em unknown} position of a stationary target and the dynamics of the robot is given by 
\begin{equation} \label{eq1}
\begin{split}
\dot {\bm p}(t) &= v_c [\cos \theta(t),\sin \theta(t)]' , \\
\dot \theta(t) &= u(t) ,
\end{split}
\end{equation}
where $\bm p(t) \in \mathbb R^2$, $v_c$, $\theta(t)$, and $u(t)$ denote the position, constant linear speed, heading course, and angular speed of the robot, respectively. 

Under mild conditions, the objective of this work is to design a proper controller $u(t)$ by using {\em range-only} measurements 
\begin{align*}
d(t) := \Vert \bm p(t) - \bm p_o \Vert_2
\end{align*}
such that the robot eventually  encircles the target in the form of any given smooth pattern, which can be specified by a smooth reference signal  $r(t)$. That is, 
\bee\label{obj}
\lim_{t\rainfty} |d(t)-r(t)|=\lim_{t\rainfty} |\dot d(t)-\dot r(t)|=0.
\ene
If $r(t)\equiv r_c$ is a positive constant, the desired pattern of the robot becomes an exact circle with the stationary target as its center. This is the celebrated circumnavigation problem in \citet{Cao2015UAV,Matveev2011Range} and see Fig. \ref{fig_encir} for illustration. However, it is confirmed via simulations in Section \ref{sec5} that their controllers cannot be directly extended to the case of time-varying $r(t)$, which is the focus of this paper. Since the controller of this work only relies on range measurements between the robot and target, it is particularly useful in the GPS-denied environment and also substantially different from \citet{Dobrokhodov2008Vision,Zhang2010Vision,Deghat2014Localization,Xiao2017Target}, all of which need the GPS state information of the robot.

\section{Controller Design}
\label{sec3}
In this section, the controller is designed by using range-only  measurements such that the distance  $d(t)$ to the target is able to track a smooth reference signal $r(t)$ in the sense of  \eqref{obj}. Our main idea is to exploit the advantage of the backstepping method, which is substantially different from \citet{Cao2015UAV,Matveev2011Range,Xiao2017Target} and they can only solve the circumnavigation problem. 

\subsection{Range-only based controller for encircling}\label{subsec_roc}

\begin{figure}
	\center
	\includegraphics[width=0.6\linewidth]{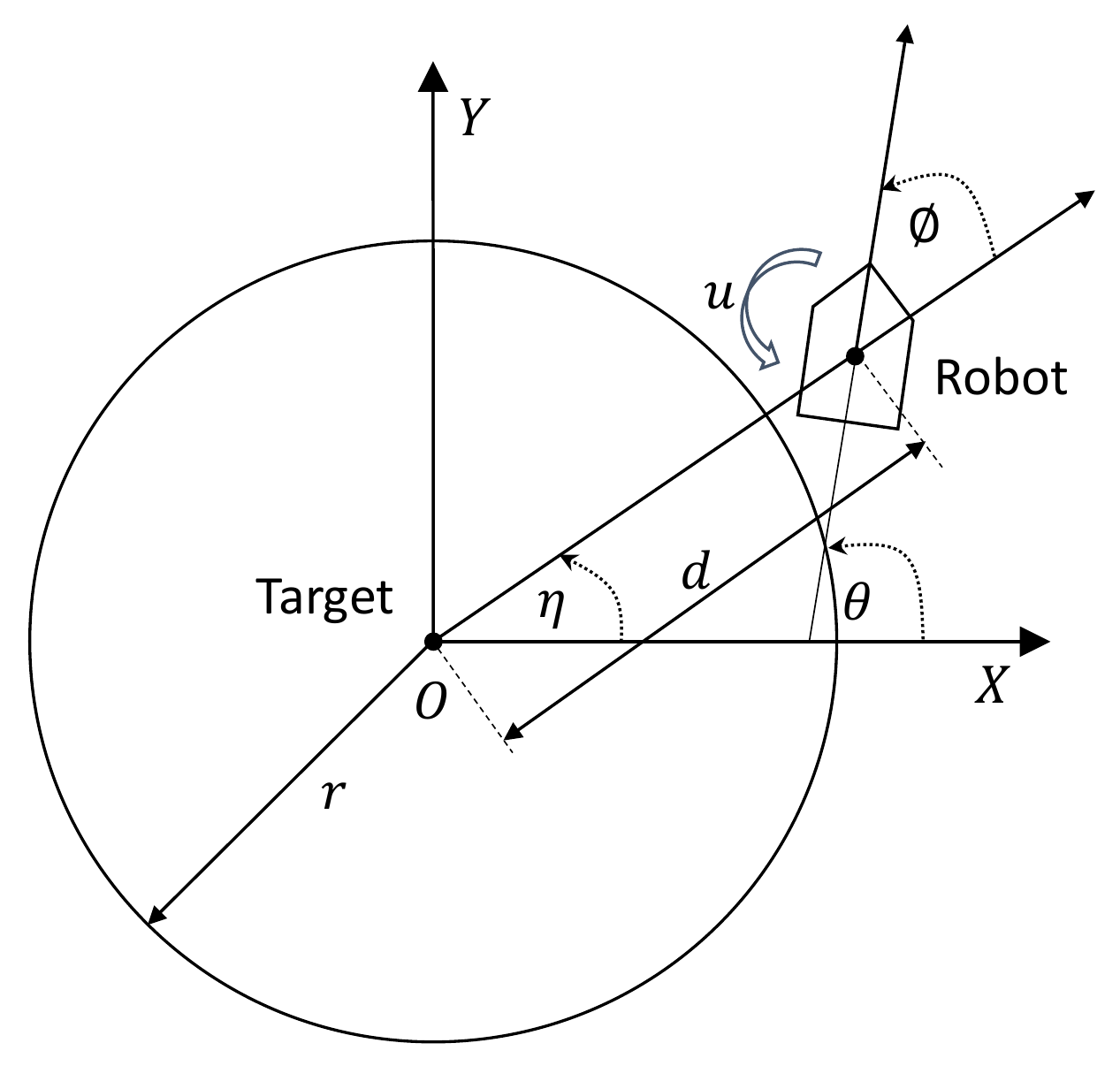}
	\caption{Target encirclement by a mobile robot.}
	 \label{fig_encir}
\end{figure}

Let $\phi(t) \in (-\pi,\pi]$ be the angle formed by the direction from the target $O$ to the robot and  the heading direction of  the robot, see Fig.~\ref{fig_encir}. By convention, the counterclockwise direction is set to be positive. Then, the robot dynamics in (\ref{eq1}) can be transformed into
\begin{equation} \label{eq5}
	\begin{split}
		\dot d(t) &= v_c \cos \phi(t),  \\
		\dot\phi(t) &= u(t) - \frac{v_c}{d(t)} \sin \phi(t). \\
	\end{split}
\end{equation}  
From Fig.~\ref{fig_encir}, we also have the relation that
\begin{align*}
	\phi(t) = \theta(t)  - \eta(t),  
\end{align*}
where $\eta(t):=\arctan( (y(t)-y_o)/(x(t)-x_o) )$ is the angle between the direction from the target to the robot and the positive direction of $x$-axis.

%

%
%
For a given smooth reference signal $r(t)$, we exploit the backstepping method to design the following range-only based controller
\begin{align} \label{eqc}
	 u(t) =&\frac{v_c  \alpha(t)}{d(t)} + \frac{1}{v_c \alpha(t)} \times  \\
	& \left(k_1 \left ( \dot d(t) - \dot r(t) + k_2 \sat \big(\frac{d(t) -r(t)}{k_3}\big)  \right  ) - \ddot r(t)   \right) , \nonumber
\end{align}
where $k_i$, $i=1,2,3$ are positive parameters, and $$\alpha(t) =|\sin \phi(t)|= \sqrt{v_c^2 - (\dot d(t) )^2}/v_c.$$  The saturation function $\sat(\cdot)$ is defined as 
\begin{align*}
	\sat(\eta) := \begin{cases}
		\eta , & |\eta|<1, \\
		\sgn(\eta), & |\eta|\ge 1, 
	\end{cases}
\end{align*}
where $\sgn(\cdot)$ is the standard signum function, and is utilized to ensure that all parameters work for any initial state of the robot. Moreover, we set $d(t) = \varepsilon_1$ if $d(t) \le \varepsilon_1$, and $\alpha(t) = \varepsilon_2$ if $\alpha(t) \le \varepsilon_2$, where $ \varepsilon_i, i=1,2$  are positive constant. Although it is extremely difficult to establish an explicit dependence, the tracking accuracy is positively related to  $k_1$, and the convergence speed is mainly  determined by  $k_2$ and $k_3$.

 Moreover,  one may employ a washout filter to track the derivative of $d(t)$. That is, $\dot d(t)$ in \eqref{eqc} is replaced by $\xi(t)$, whose Laplace transform is given by 
\begin{align} \label{eq4}
	\xi(s) = \frac{hs}{s+h}d(s),
\end{align}
where $s$ is the Laplace operator, $h>0$ is the filter parameter, and $d(s)$ is the Laplace transform of the range-only input. Here the washout filter is used to damp the DC component of the range measurement $d(t)$ \citep{Lin20163}. 
Noting that $\xi(t)=\dot d(t)$ if there is no measurement noise, and otherwise $\xi(t)$ is a filtered version of $\dot d(t)$. A similar idea of (\ref{eq4}) has also been adopted in \citet{Guler2015Range}.


\subsection{Interpreting  \eqref{eqc} from the backstepping control} \label{subsec3}
Taking the derivative on both side of \eqref{eq5}, we obtain that 
\begin{align} \label{eq23}
	\ddot d(t)  = -\dot\phi(t) v_c \sin \phi(t).
\end{align}
Consider the following state vector of the robot
 \bee\label{statedef}
\bm x(t) := [x_1(t),x_2(t)]' = [d(t), \dot d(t)]'.
\ene
Then, it follows from \eqref{eq23} that
\begin{equation} \label{eq58}
	\begin{split}
		\dot x_1(t) &= x_2(t),  \\
		\dot x_2(t) &= f(\bm x(t)) + g(\bm x(t)) u(t), \\
	\end{split}
\end{equation} 
where the functions $f$ and $g$ are given by
\begin{equation} \label{eq59}
	\begin{split}
		f(\bm x(t)) &= \frac{v_c^2}{x_1(t)} \sin ^2 \phi(t),  \\
		g(\bm x(t)) &= -v_c \sin\phi(t). \\
	\end{split}
\end{equation} 
Define an error vector $\bm e(t) := [e_1(t),e_2(t)]'$ by
\begin{equation} 
	\begin{split}
		e_1(t) :=& x_1(t) - r(t), \\
		e_2(t) := & x_2(t) - s(t), \\
	\end{split}
\end{equation} 
where $s(t)$  is a virtual guidance command and is defined as
\begin{align} \label{eq29}
	s(t) :=  - c_1 e_1(t) + \dot r(t) .
\end{align}	
Here $c_1>0$ is a control parameter, and $ \dot r(t) $ is the derivative of the desired command $r(t)$.

Consider the Lyapunov function candidate as
\begin{align*}
	V_1(e_1) = \frac{1}{2} e_1^2(t).
\end{align*}
Taking the derivative of $V_1(e_1)$, we have that
\begin{align*}
	\dot V_1(e_1) &= e_1 (t)\dot e_1(t)\\
	& = -c_1 e_1(t)^2 + e_1(t)e_2(t).
\end{align*}
If $e_2(t)=0$, then $\dot V_1(e_1) \le 0$ and $x_2(t) = s(t)$. Moreover, 
$$
	\dot e_1(t) = - c_1 e_1(t).
$$
In this case, $e_1(t)$ exponentially converges to $0$ as $t$ goes to infinity.  Thus, it is sufficient to design a proper controller $u(t)$ to asymptotically drive $e_2(t)$  to zero. A natural idea is to use a backstepping controller \citep{Khalil2002Nonlinear,ZHONG1997Tracking}, i.e., 
\begin{align} \label{eq48}
	u(t) =  -\frac{c_2 e_2(t) + e_1(t) +f(\bm x(t)) -\dot s(t) }{g(\bm x(t))} ,
\end{align}
where $c_2>0$ is a positive parameter, and $\dot s(t)$ is the derivative of the virtual command $s(t)$. 

To validate the controller in \eqref{eq48}, consider the Lyapunov function candidate as 
\begin{align*}
	V_2(\bm e) = V_1(e_1) + \frac{1}{2} e_2^2(t).
\end{align*}
Taking derivative of $V_2(\bm e)$ along with \eqref{eq58} leads to that 
\begin{align*}
	\dot V_2(\bm e) &=  e_1(t) \dot e_1(t)+ e_2(t)\dot e_2(t) \\
	&= e_1(t) \dot e_1(t)  + e_2(t) \big( f(\bm x(t)) + g(\bm x(t)) u(t)  - \dot s(t) \big) \\
	&= -c_1 e_1^2(t)  -c_2 e_2^2(t)\\
	&  \le 0 .
\end{align*}
Thus, $V_2(\bm e)$ satisfies the following three conditions 
\begin{itemize}
	\item Nonnegative: $V_2( \bm e)\ge 0$, and $V_2( \bm e) = 0$ if and only if $\bm e(t)= \bm0$.
	\item Strictly decreasing: $\dot V_2(\bm e) <0$, $\forall \bm e(t) \neq \bm 0$.
	\item Radially unbounded: $V_2(\bm e) \rightarrow \infty$, as $\Vert \bm e(t) \Vert _2 \rightarrow \infty$.
\end{itemize}
By Theorem 4.1 in \citet{Khalil2002Nonlinear}, the closed-loop system in  (\ref{eq58}) is asymptotically stable, i.e., $d(t)\rightarrow r(t)$ and $\dot d(t)\rightarrow \dot r(t)$ as $t\rightarrow \infty$. 

However, the backstepping controller in (\ref{eq48}) further uses the information of the angle $\phi(t)$. To solve it, we note from the dynamics in (\ref{eq5}) that 
$\cos \phi(t) =x_2(t)/v_c$. Since $\sin \phi(t) = \pm \sqrt{1-\cos^2\phi(t)}$ and the sign of $\sin \phi(t)$ is unknown, we simply use $\alpha(t)=|\sin\phi(t)|$ to replace it and the controller in (\ref{eq48}) is modified as 
\begin{align}  \label{eq17}
u(t) 
&= - \frac{v_c \alpha(t)}{x_1(t)}  - \frac{1}{v_c \alpha(t)} \big(c_2 e_2(t) + e_1(t)  -\dot s(t) \big)  \nonumber\\
&= -{v_c\alpha(t)}/{x_1(t)}  - {1}/(v_c \alpha(t)) \times \nonumber\\
 &~~~~(c_2 e_2(t) + e_1(t)  + c_1 x_2(t) - c_1 \dot r(t)- \ddot r(t))  \nonumber\\
&=  -{v_c\alpha(t)}/{x_1(t)}- {1}/(v_c \alpha(t)) \times \\
&~~~~{ \left (k_1 \left(   x_2(t) - \dot r(t)  +\displaystyle\frac{x_1(t) -r(t)}{k_3}\right)     - \ddot r(t)\right) },  \nonumber
\end{align}
where $k_1 = c_1 + c_2$, and $k_3= (c_1 +c_2 )/(c_1 c_2 + 1)$.

When $\phi(t) \in [0,\pi]$, it is clear that the controller in (\ref{eq17}) is equivalent to that in (\ref{eq48}). Thus, if there is a finite $t_1 \ge t_0$ such that $\phi(t) \in [0,\pi]$, $\forall t\ge t_1$ for any initial state $\phi(t_0)$, the controller in (\ref{eq17}) can guarantee the global convergence.  

To ensure the existence of such a finite $t_1$, the parameter $k_3$ usually depends on the initial state of $x_1(t)$. Note that $|x_2(t)| \le v_c $, we adopt a saturation function to handle this problem. Then,  the controller in (\ref{eq17}) is further modified as \eqref{eqc}.

In the sequel, we shall prove that the controller in (\ref{eqc}) indeed drives the robot to encircle the target in the sense of (\ref{obj}). 

\subsection{Comparison with the literature } \label{subsec5}
For the circumnavigation problem, i.e., $r(t) \equiv r_c$, the proposed controller in (\ref{eqc}) naturally reduces as 
\begin{align} \label{eq22}
	u(t) =\frac{v_c \alpha(t)}{d(t)}  +   
	\frac{k_1}{v_c \alpha(t)}{ \left (  \dot d(t) + k_2 \sat (\frac{d(t) -r_c}{k_3})   \right)}.
\end{align}


When the robot is sliding on the desired orbit, i.e., $d(t)=r_c$ and $\dot d(t)=0$, the control output is exact $v_c/r_c$, which implies that there is no steady-state error. In contrast, the sliding mode controller in \citet{Matveev2011Range} cannot achieve zero steady-state error. 
Since this issue, \citet{Cao2015UAV} and \citet{Zhang2017Unmanned} use switching controllers along with the distance error $d(t)-r_c$. The idea of the geometrical method in \citet{Cao2015UAV} is  to drive the robot towards the tangent point of an auxiliary circle, and there is no control when the robot enters the auxiliary circle, which may result in large overshoot.  

In terms of the stability and convergence, our controller in \eqref{eq22} can ensure global convergence, and the control parameters are independent of the initial distance. In comparison, 
the sliding mode approach in \citet{Matveev2011Range} requires that the initial distance to the target is lager enough than the desired radius.  The control parameters  in \citet{Milutinovi2017Coordinate} are determined by solving a linear quadratic regulator (LQR) problem and depends on the initial distance.

Note that all methods mentioned above are concerned with the problem of target circumnavigation. Their methods cannot be directly extended to the case of smooth time-varying reference commands as confirmed in Section \ref{sec5}.

\section{Target Encirclement with a Constant Reference Distance} \label{sec4}
As mentioned before,  circumnavigation is a special case of target encirclement of this work, and is shown to be achieved under controller in \eqref{eqc} by setting $r(t)\equiv r_c$.

\subsection{Stability and convergence} \label{sub1}

\begin{prop} \label{prop2} Consider the encirclement system in (\ref{eq58}) under the range-only based controller in (\ref{eqc}). Let $r(t)\equiv r_c$ and $\bm x_e := [r_c, 0]'$.  If the controller parameters are selected to satisfy that
\begin{align}  \label{eqcon}
	0<k_2 < v_c,  ~\text{and}~k_3 = {r_c} 
\end{align} 
 there exists a finite $t_1>t_0$ such that 
 \begin{align*}
 \twon{\bm x(t)-\bm x_e}\le C\twon{\bm x(t_1)-\bm x_e} \exp\left(-\rho (t-t_1)\right), \forall t>t_1
 \end{align*}
 where $\bm x(t)$ is defined in \eqref{statedef}, $\rho$ and $C$ are two positive constants.  
\end{prop}

Thus, a proper set of parameters in  (\ref{eqcon})  ensures that the robot completes the circumnavigation task for any initial state and the convergence is exponentially fast. 

If $\bm x(t) =\bm x_e$, the controller in (\ref{eqc}) reduces to $u(t)=v_c/r_c$, which together with \eqref{eq5} implies that there is no steady-state error. 
In contrast,  the sliding mode approach  in \citet{Matveev2011Range} cannot achieve zero steady-state error.


\subsection{Proof of Proposition \ref{prop2}} \label{sub2}

If $\phi(t) \in [0,\pi]$, the controller in (\ref{eq17}) is equivalent to that in (\ref{eq48}), the effectiveness of which has been proved in Section \ref{subsec3}. Thus, we first show that there must exist a finite time instant $t_1\ge t_0$ such that $\phi(t) \in [0,\pi], \forall t\ge t_1$ for any initial state, see Lemma \ref{lemma3}. Then, the closed-loop system in (\ref{eq58}) under (\ref{eq22}) is shown to be asymptotically stable in Lemma \ref{lemma4} and  exponentially stable in Lemma \ref{lemma5}.

\begin{lem} \label{lemma3} 
	Under the conditions in Proposition \ref{prop2}, there must exist a finite time instant $t_1\ge t_0$ such that $\phi(t) \in [0,\pi]$, $\forall t \ge t_1$, for any initial state $\phi(t_0) \in (-\pi,\pi]$.
\end{lem}
\begin{pf} 	We prove that (a) $\phi(t) \in [0,\pi]$ for any $t\ge t_0$ if $\phi(t_0) \in [0,\pi]$ and (b) there exists a $t_1> t_0$ such that $\phi(t_1) \in [0,\pi]$ for any $\phi(t_0) \in (-\pi,0)$.
	 
	 Combining (\ref{eq5}) with  (\ref{eq22}), we obtain that
	 	\begin{align}\label{eqphi}
	 		\dot \phi(t)& = \frac{v_c \alpha(t)}{d(t)}  +   
	 		\frac{k_1}{v_c \alpha(t)}{ \big(  \dot d(t) + k_2 \sat (\frac{d(t) -r_c}{k_3})   \big)}\nonumber\\  
	 		                     &- \frac{v_c}{d(t)} \sin \phi(t).
	 		\end{align}
		
		To prove part (a), we just need to show that $\dot \phi(t) > 0$ if $\phi(t) = 0$, and $\dot \phi(t) <  0$ if $\phi(t) = \pi$. If $\phi(t) = 0$, it follows from (\ref{eqphi}) that 
		\begin{align*} 
			\dot \phi(t)& =   (v_c \varepsilon_2)^{-1} k_1 \left (  v_c + k_2 \sat (\frac{d(t) -r_c}{k_3})   \right)\\
			& \ge (v_c \varepsilon_2)^{-1} k_1 (  v_c - k_2 ) > 0,
		\end{align*}
		where the last inequality uses the fact that $ {v_c} >  k_2  $. 
Similarly, $\phi(t) = \pi$ leads to that
		\begin{align} \label{eq36}
			\dot \phi(t)  &= (v_c \varepsilon_2)^{-1} k_1 \left (  -v_c + k_2 \sat (\frac{d(t) -r_c}{k_3})   \right)\nonumber \\
			&\le (v_c \varepsilon_2)^{-1} k_1  (  -v_c + k_2 ) < 0.
		\end{align}
	
	To prove part (b), four cases in Fig. \ref{fig16} are considered.	
	
	\begin{figure}[t!]
		\centerline{\includegraphics[width=0.9\linewidth]{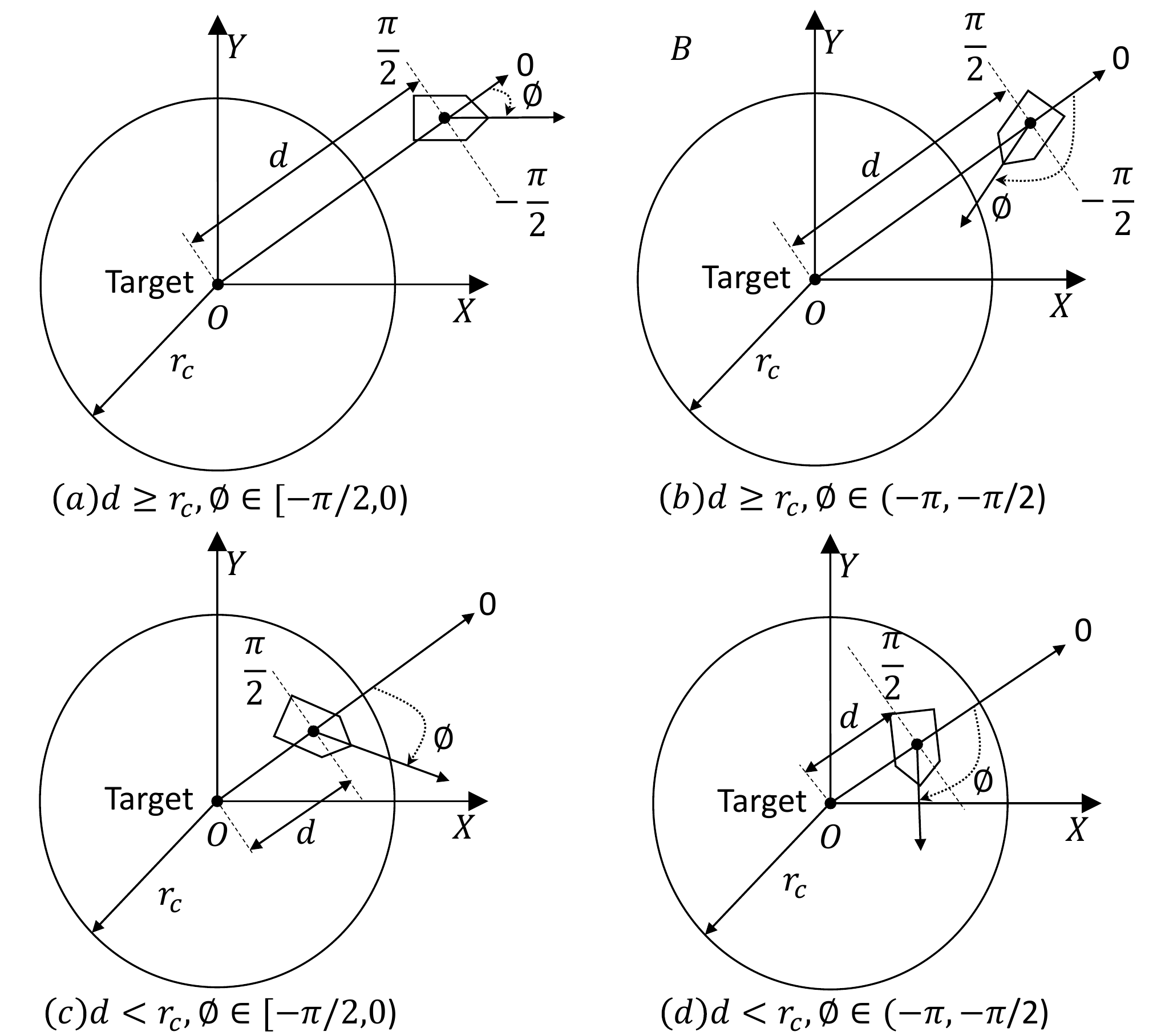}}
		\caption{States of the robot.}
		\label{fig16}
	\end{figure}

For the case in Fig. \ref{fig16}(a), i.e., $d(t_0) \in [r_c,\infty)$ and $\phi(t_0) \in [-\pi/2,0)$,  it follows from \eqref{eqc} and (\ref{eqphi}) that $\dot d(t_0) \ge 0$ and $\dot \phi(t_0) >0$. Thus, there must exist a $t_*>t_0$ such that $0> \phi(t_*) >- \pi/2$ and $\dot d(t_*)>0$. This implies that $d(t_*)>r_c$, and 
\begin{align*}
	\dot \phi(t_*)& = - \frac{k_1}{v_c \sin \phi(t_*)}{ \left (  \dot d(t_*) + k_2 \sat (\frac{d(t_*) -r_c}{k_3})   \right)}\\
                        &~~~~	-\frac{2 v_c \sin \phi(t_*) }{d(t_*)}\\
                        & > -{k_1}{\cot \phi(t_*) }  \\
                        &>0.
\end{align*}
Thus, $\phi(t)$ will monotonically increase until $\phi(t_1)\ge 0$ at some finite time instant $t_1$.

		For the case in Fig.~\ref{fig16}(b), i.e., $d(t_0) \in [r_c,\infty)$ and $\phi(t_0) \in (-\pi,-\pi/2)$, it follows from  (\ref{eqphi}) and (\ref{eq36}) that 
		\begin{align*}
		 \begin{cases}
		\dot \phi(t)>0 ,& \text{if}~ \phi(t) = -\pi/2, \\
		\dot \phi(t) <0,& \text{if}~ \phi(t)=-\pi.
		\end{cases} 
		\end{align*}
		 In this case, there are three possible results after some finite time $\Delta>0$: (i) $\phi(t_0 + \Delta) <-\pi $  and $d(t_0+\Delta)\ge r_c$, then part (b) is finished; (ii)  $\phi(t_0+\Delta) >-\pi/2$ and $d(t_0+\Delta)\ge r_c$, which is the case in Fig.~\ref{fig16}(a);  (iii)  $d(t_0+\Delta)< r_c$ which is to be shown for the cases in Fig.~\ref{fig16}(c) and (d).  
	
For $d(t_0) \in (0,r_c)$, it contains both cases in Fig.~\ref{fig16}(c) and (d). When $\phi(t)=-\pi/2$, it follows from (\ref{eqphi}) that
\begin{align} \label{eq35}
	\dot \phi(t) &=\frac{-2v_c}{d(t)} - \frac{k_1 k_2}{r_c v_c } ({d(t) -r_c})  \\
	                & = \frac{ -d^2(t) + r_c d(t) - {2r_c v_c^2}/{(k_1k_2)}}{ {r_c v_c d(t)}/{(k_1k_2)}} .
\end{align}
Note that the numerator in the last equality is quadratic in $d(t)$.
If $k_1k_2<8 v_c^2$, then it can be easily verified that 
\begin{align*}
r_c^2 - 4 \times \frac{2r_c v_c^2}{k_1k_2} <0
\end{align*} 
and $\dot \phi(t)<0$ for $\phi(t)=-\pi/2$ and any $d(t) \in (0,r_c)$. 
If $k_1k_2\ge 8 v_c^2$, there exists an  equilibrium $\tilde{\bm y}_e:=[d_*, -\pi/2]'$ such that $\dot \phi(t)=0$, where $d_* \in (0,r_c)$. However, this equilibrium $\tilde{\bm y}_e$ is unstable.
To prove it, we define $\bm y(t) := [d(t),\phi(t)]'$ and linearize the closed-loop system in \eqref{eq5} around $\tilde{\bm y}_e$ as follows
\begin{align*}
\dot {\bm y}(t) = F (\bm y(t) -\tilde{\bm y}_e )
\end{align*} 
where the Jacobian matrix $F$ is given by
\begin{align*}
F = \bmatri 0 & v_c \\ \displaystyle \frac{k_1 k_2}{r_c v_c} -\frac{2 v_c }{d_*^2}  & k_1  \ematri.
\end{align*}
It is clear that at least one eigenvalue of $F$ has positive real part, i.e., the equilibrium $\tilde{\bm y}_e$ is unstable.
Thus, $\phi(t)$ cannot converge to $-\pi/2$ for any $d(t) \in (0,r_c)$.
%
%

Suppose that $\phi(t)$ can stay in $(-\pi/2, 0)$ for all time, then we have $\dot d(t) > 0$ in light of (\ref{eq5}). Together with the fact that $\phi(t)$ cannot converge to $-\pi/2$ for any $d(t) \in (0,r_c)$, there must exist some finite time $\Delta > 0$ such that $d(t_0+\Delta )\ge r_c$, which is the case in Fig.~\ref{fig16}(a).


Similarly, $d(t)>0$ implies that $\phi(t)$ cannot stay in $(-\pi, -\pi/2)$ for all time. 

Consequently, there must exist a finite $t_1$ such that $\phi(t_1)\in [0,\pi]$ for any $\phi(t_0) \in (-\pi,0)$. 	 \qed

	
	
\end{pf} 

In virtue of Lemma \ref{lemma3},  the controller in (\ref{eq17}) is eventually equivalent to that in (\ref{eq48}), whose effectiveness has been shown in Section \ref{subsec3}.


\begin{lem} \label{lemma4} Under the conditions in Proposition \ref{prop2},  the closed-loop system in (\ref{eq58}) is asymptotically stable.    
\end{lem}

\begin{pf} In view of Lemma \ref{lemma3}, it holds that $\phi(t) \in [0,\pi]$ for any $t\ge t_1$. By  (\ref{eqphi}), we obtain that
	 \begin{align}\label{eqphi1}
	 \dot \phi(t) = 
	 \frac{k_1}{v_c \sin \phi(t)}{ \left (  \dot d(t) + k_2 \sat (\frac{d(t) -r_c}{k_3})   \right)}.
	 \end{align}

	Consider the Lyapunov function candidate as 
	\begin{align*}
		V_3(\bm x) =  k_1 k_2 \int_{r_c}^{x_1(t)} {\sat (\frac{\tau -r_c}{k_3})}\text{d}\tau + \frac{1}{2} x_2^2(t).
	\end{align*}
	Taking the time derivative of $V_3(\bm x )$ along with (\ref{eq23}), (\ref{eq58}) and (\ref{eqphi1}) leads to that 
	\begin{align*}
		\dot V_3(\bm x ) & =  k_1 k_2 {\sat (\frac{x_1(t) -r_c}{k_3})} x_2(t)   +  x_2(t) \dot x_2(t)   \\ 
		          	                                       & =  k_1 k_2 {\sat (\frac{x_1(t) -r_c}{k_3})} x_2(t)  \\
		          	                                            &~~~~ -  x_2(t)   \left(  k_1 x_2(t) +k_1  k_2 \sat (\frac{x_1(t) -r_c}{k_3})   \right)      \\
		&= -k_1 x_2^2(t) \\
		& \le 0.
	\end{align*}
	
	However, $\dot V_3(\bm x)$ is not always negative definite, e.g.,  $\dot V_3(\bm x) =0$ for $x_2(t) = 0$ and any $x_1(t)>0$.  
	Let $\mathcal S:=\{\bm x  | \dot V_3(\bm x) =0 \}$. For any ${\bm x}_0\in \mathcal S$ but ${\bm x}_0\neq \bm x_e$, it holds that 
	\begin{align*}
		\dot  x_2(t) |_{\bm x(t) = {\bm x}_0}=   k_1  k_2\sat\big(  \frac{x_1(t)-r_c}{k_3} \big)   \neq 0.
	\end{align*}
	Thus, $x_2(t)$ cannot maintain the state $\bm x(t)=\bm x_o$, i.e., ${\bm x}_0$ cannot stay identically in $\mathcal S$. 
	
	
	Moreover, $V_3(\bm x)$ is nonnegative and radially unbounded. By the LaSalle's invariance theorem  \citep[Corollary 4.2]{Khalil2002Nonlinear}, $\bm x_e=[r_c,0]'$ is an asymptotically stable equilibrium point of the closed-loop system in (\ref{eq58}). \qed
	
\end{pf}

If the initial position of the robot is far away from the target, i.e., $d(t_0) -r_c \gg k_3$ and $\phi(t_0)\in[0,\pi]$.  
It follows from (\ref{eqphi1}) that 
\begin{align} \label{eq37}
	 \dot \phi(t_0) = 
	 \frac{k_1}{v_c \sin \phi(t_0)}{ \left ( v_c \cos \phi(t_0) + k_2  \right)}.
	 \end{align}

For simplicity,  we define $\varphi:= \cos^{-1}(-k_2/v_c).$ 
If $\phi(t_0)\in[0,\varphi)$, then $\dot \phi(t_0) >0 $ according to (\ref{eq37}). Hence, $\phi(t)$ monotonically increases until $\phi(t_*)=\varphi$ at some finite time instant $t_*>t_0$. Similarly, if $\phi(t_0)\in(\varphi,\pi]$, then $\phi(t)$ monotonically decreases until $\phi(t_*)=\varphi$. Once $\phi(t)=\varphi$, it follows from (\ref{eq37}) that $\dot \phi(t) =0$, i.e., $\dot d(t) + k_2=0$. Consequently, $d(t)$ converges to $r_c$ at the speed $\dot d(t) = -k_2$ until $d(t_2)-r_c=k_3$, where  $t_2>t_1$ is finite. The above statement is to be validated in Subsection \ref{subsec1}. 


   Lemma \ref{lemma5} further proves that the proposed controller in (\ref{eq22}) is able to ensure that the convergence is exponentially fast.


\begin{lem} \label{lemma5} Under the conditions in Proposition \ref{prop2}, there must exist a finite $t_1\ge t_0$ such that
	\begin{align*}
	\twon{\bm x(t)-\bm x_e}\le C\twon{\bm x(t_1)-\bm x_e} \exp\left(-\rho (t-t_1)\right), \forall t>t_1
	\end{align*}
	where $\rho$ and $C$ are two positive constants.	
\end{lem}
\begin{pf}
	In view of Lemma \ref{lemma4}, there must exist a finite $t_1\ge t_0$ such that $|x_1(t_1)-r_c|<k_3$ and $\phi(t_1) \in [0,\pi]$. Thus, for any $t> t_1$, it follows from (\ref{eq58}) and (\ref{eq22}) that
	\begin{align} \label{eq34}
		\dot {\bm x}(t) = A (\bm x(t) - \bm x_e),
	\end{align}
	where 
	\begin{align} \label{eq28}
		A  =   \bmatri 0  &  1 \\
		-{k_1 k_2}/{k_3} & -{k_1 } \ematri  .              
	\end{align}

	
	One can easily verify that both the eigenvalues of $A$ have negative real part, i.e., $A$ is Hurwitz. 

In addition, it follows from (\ref{eq34}) that the trajectory of this system after $t_1$ must satisfy 
	\begin{align*}
	\bm x(t) -\bm x_e = Q\exp\left(\Lambda (t-t_1)\right)Q^{-1} (\bm x(t_1)-\bm x_e), \forall t> t_1,
	\end{align*}
	where $A=Q\Lambda Q^{-1}$, $\Lambda= \diag( \lambda_1, \lambda_2 )$, and $\lambda_i$, $i=1,2$ is the eigenvalue of $A$. Finally, it holds that
	\begin{align*}
	\twon{\bm x(t) -\bm x_e } &= \twon{ Q\exp(\Lambda (t-t_1))Q^{-1} (\bm x(t_1)-\bm x_e)} \\
	&\le C\twon{\bm x(t_1)-\bm x_e} \exp(- \rho (t-t_1))
	\end{align*}
	where $C=\twon{Q}\twon{Q^{-1}}$, 
	\begin{align*}
	\rho =
	\begin{cases}
    (k_1 - \sqrt{\Delta})/2 &~,\text{if}~ \Delta >0, \\
    k_1/2&, ~\text{if}~\Delta \le 0,
	\end{cases}
	\end{align*}
	and
	$\Delta =k_1 ^2 - 4 \times {k_1 k_2}/{k_3}$.
	 \qed

\end{pf}	 

{\em  Proof of Proposition \ref{prop2}}. 
In view of Lemma \ref{lemma3}, there must exist a finite $t_1\ge t_0$ such that $\phi(t) \in [0,\pi]$, $\forall t \ge t_1$, for any $\phi(t_0)\in (-\pi,\pi]$. Once $\phi(t)$ enters the region $[0,\pi]$, the closed-loop system in (\ref{eq58}) asymptotically converges to the equilibrium $\bm x_e=[r_c,0]'$ by Lemma \ref{lemma4}. Thus $\bm x_e$ is a global stable equilibrium for any initial states. Furthermore, if $|x_1(t)-r_c|<k_3$ and $x_2(t)(x_1(t)-r_c)$, the convergence is exponentially fast based on Lemma \ref{lemma5}.


\section{Target Encirclement with Smooth Patterns}  \label{sec7}   

In this section, we show that the controller in (\ref{eqc}) is further able to follow smooth time-varying $r(t)$. For example, the command in (\ref{obj}) is generated by a sine function. In comparison, either the sliding mode approach in \citet{Matveev2011Range} or the geometrical method in \citet{Cao2015UAV} is unable to complete this task.

\begin{prop} \label{prop3} Consider the encirclement system in (\ref{eq58}) under the range-only based controller in (\ref{eqc}). Suppose that the command $r(t)$ is twice continuous differentiable, and $\dot r(t)$, $\ddot r(t)$ denote the first and second order derivatives of $r(t)$, respectively. If  $|\dot r(t)|\le r_v$, $ |\ddot r(t)| \le r_a$, and the controller parameters are selected to satisfy that 
\begin{align} \label{eq3}
	k_1 > \frac{k_2}{k_3},~ k_1(v_c-k_2-r_v) > r_a,~\text{and}~ k_1(v_c^2 -r_v^2) > r_v r_a,
\end{align}
and $\phi(t_0) \in [0,\pi]$,
then there exists a finite $t_1>t_0$ such that 
\begin{align*}
\twon{\bm z(t)}\le C\twon{\bm z(t_1)} \exp\left(-\rho (t-t_1)\right), \forall t>t_1
\end{align*}
where $\bm z(t):=\bm x(t) -\bm x_e(t)$, $\bm x_e(t): =[r(t),\dot r(t)]'$, $\rho$ and $C$ are two positive constants. 
\end{prop}

\begin{pf}
	Firstly, we show that $\phi(t) \in [0,\pi]$ holds for all $ t \ge t_0$.
	
	Submitting (\ref{eqc}) into (\ref{eq5}) yields that 
	\begin{align} \label{eq27}
	\dot \phi(t) &= \frac{v_c  \alpha(t)}{d(t)} - \frac{v_c}{d(t)} \sin \phi(t) + \frac{1}{v_c \alpha(t)} \times  \\
	& ~~~~\left(k_1 \left ( \dot d(t) - \dot r(t) + k_2 \sat \big(\frac{d(t) -r(t)}{k_3}\big)  \right  ) - \ddot r(t)   \right)  \nonumber.
	\end{align}
	
	If $\phi(t) =0$, it follows from (\ref{eq3}) and (\ref{eq27})  that
	\begin{align*}
		\dot \phi(t) &=   
		(v_c \varepsilon_2)^{-1} \left(k_1 (  v_c + k_2 \sat (\frac{e_1(t)}{k_3})   - \dot r(t) ) - \ddot r(t)   \right)  \\
		&\ge (v_c \varepsilon_2)^{-1} \left(k_1 (  v_c - k_2    - r_v ) - r_a   \right) > 0. 
	\end{align*}
	Similarly, if $\phi(t) =\pi$, then
	\begin{align*}
		\dot \phi(t) &=   
		(v_c \varepsilon_2)^{-1} \left(k_1 (  -v_c + k_2 \sat (\frac{e_1(t)}{k_3})   - \dot r(t) ) - \ddot r(t)   \right)  \\
		&\le (v_c \varepsilon_2)^{-1} \left(k_1 ( - v_c + k_2   + r_v ) + r_a   \right) < 0.
	\end{align*}

	Hence,  $\phi(t) \in [0,\pi]$ holds for any $t \ge t_0$.
	
	Let 
	\begin{align} \label{eq44}
		s(t) &:= -k_2\sat (\frac{e_1(t)}{k_3}) + \dot r(t),\nonumber\\
		e_1(1)&:= d(t) -r(t),\\
		{e}_2(t) &: = \dot d(t) -s(t). \nonumber
	\end{align} 
	Then, the proposed controller in (\ref{eqc}) can be rewritten as  
	\begin{align*}
		u(t) = {v_c \alpha(t)}/{d(t)}  + (v_c \alpha(t) )^{-1} (k_1 {e}_2(t) -\ddot r(t)).
	\end{align*}
	Consider the Lyapunov function candidate as 
	\begin{align*}
		& V_4(\bm e)= \frac{1}{2} {e}_2^2(t) /2+\\
		&~~~~\frac{k_2^3}{k_3} \left(\int_{r(t)}^{d(t)} {\sat (\frac{\tau -r(t)}{k_3})}\text{d}\tau + \int^{r(t)}_{d(t)} {\sat (\frac{r(t) -\tau }{k_3})}\text{d}\tau   \right).
	\end{align*}
	Taking the time derivative of $ V_4(\bm e)$, we obtain that
	\begin{align*}
		&\dot  V_4(\bm e) = \frac{k_2^2}{k_3}\sat(\frac{e_1(t)}{k_3}) \dot e_1(t) + {e}_2(t) \dot {e}_2(t) \\
		&~~~~= \frac{k_2^2}{k_3}\sat(\frac{e_1(t)}{k_3}) \dot e_1(t) + {e}_2(t) (-k_1 {e}_2(t) +\ddot r(t) - \dot s(t)).  
	\end{align*}
	
	If $|e_1(t)| < k_3$, it follows from (\ref{eq44}) that
	\begin{align*}
		\dot s(t) = -{k_2 \dot e_1(t)}/{k_3}  + \ddot r(t)
	\end{align*}
	which together with  (\ref{eq3}) implies that
	\begin{align*}
		\dot  V_4(\bm e) 
		&= -\frac{k_2^3}{k_3^3}{e_1^2(t)} -(k_1 -\frac{k_2}{k_3}){e}_2^2(t)    \\
		&\le  0.    
	\end{align*}	
	If $|e_1(t)| \ge k_3$, we have $\dot s(t) =  \ddot r(t)$, and 
	\begin{align} \label{eq39}
		\dot  V_4(\bm e)  
		=-k_1 {e}_2^2(t) + \frac{k_2^2}{k_3}\sgn\left(e_1(t) \right) {e}_2(t) - \frac{k_2^3}{k_3}.
	\end{align}
	Since the right hand side of (\ref{eq39}) is quadratic in ${e}_2(t)$,  it can be easily verified that
	\begin{align*}
	 \left(\frac{k_2^2}{k_3}\sgn(e_1(t)) \right)^2 - 4 \times (-k_1)(- \frac{k_2^3}{k_3}) 
&= \frac{k_2^3}{k_3}( \frac{k_2}{k_3} - 4k_1 )  \\
		&< 0.
	\end{align*}
     Hence, it holds that
	 $\dot  V_4(\bm e) \le 0.$
	
	 One can easily verify that the Lyapunov function candidate $ V_4(\bm e)$ is nonnegative, strictly decreasing, and radially unbounded. Thus, ${\bm e} (t)$ asymptotically converges to $\bm 0$ as $t\rightarrow \infty$, which implies that the states $\bm x(t)= [d(t), \dot d(t)]'$ asymptotically converge to $\bm x_e(t) = [r(t), \dot r(t)]'$. 
	 
	 Thus, there must exist a finite $t_1\ge t_0$ such that $|z_1(t_1)|<k_3$. For any $t>t_1$, it follows from
	  (\ref{eqc}) and (\ref{eq58}) that
	\begin{align}\label{eq31}
		\ddot d(t) = -k_1\left(  z_2(t)  + \frac{k_2}{k_3} z_1(t) \right) + \ddot r(t)
	\end{align}
 and
	\begin{align} \label{eq33}
		\dot {\bm z}(t) = F \bm z(t)
	\end{align}
	where  
	\begin{align*}
		F = \bmatri  0 & 1 \\ -k_1 k_2 /k_3 & - k_1  \ematri.
	\end{align*}
    The rest of the proof is of the same as that in Lemma \ref{lemma5}.	
	\qed

\end{pf}

\section{Simulations} \label{sec5}

Consider a robot as described in (\ref{eq1}), and set the linear speed $v_c$ as $0.5$\si{m/s}.    
For brevity, let $\bm y(t) : = [\bm p'(t),\theta(t)]'$ denote the state of the robot.

\begin{table}[t]
	\caption{Parameters of the proposed controller}
	\label{tab2}%
	\centering	
	\begin{tabular}{|c|c|c|c|c|c|c|}	
		\hline
		{Parameter}   &{$k_1$} &{$k_2$}&{$k_3$}&{$h$} &$\varepsilon_1$&$\varepsilon_2$  \\
		\hline
		{Value}         & 20          & 0.45      & 2.0      & 100 & 0.01 & 0.01     \\       
		\hline
	\end{tabular}%
\end{table}%

\subsection{Target encirclement with a constant distance} \label{subsec1}

\begin{figure}[t!]
	\centerline{\includegraphics[width=0.8\linewidth]{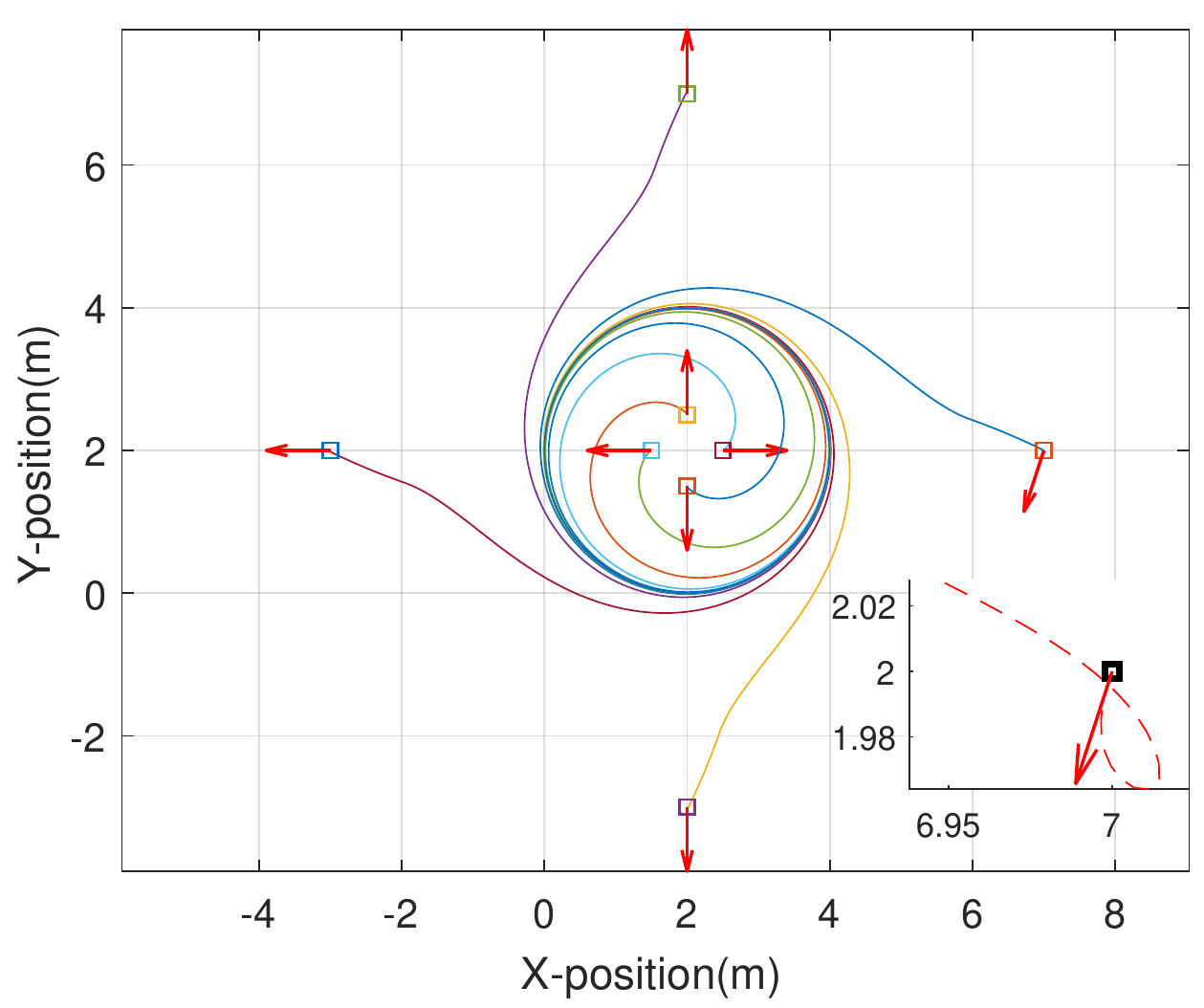}}
	\caption{Trajectories of the robot with different initial states.}
	\label{fig3}
\end{figure}

Let the target position be $\bm p_o = [2,2]'$. To test the global convergence of the proposed controller, we select eight different initial states for the robot, e.g., $\bm y(t_0)=[7,2,-3\pi/5]'$, $[2,7,\pi/2]'$, $[-3,2,\pi]'$, $[2,-3,-\pi/2]'$,   $[2.5,2,0]'$, $[2,2.5,\pi/2]'$, $[1.5,2,\pi]'$, and $[2,1.5,-\pi/2]'$, see Fig.~\ref{fig3}.
 The square and the arrow denote the initial position and initial course. The control parameters are given in Table \ref{tab2}, satisfying (\ref{eqcon}).
 From Fig.~\ref{fig3}, all trajectories of the robot form a circle centered at the target with the radius $r_c=2$. 

\begin{figure}[t!]
	\centerline{\includegraphics[width=0.8\linewidth]{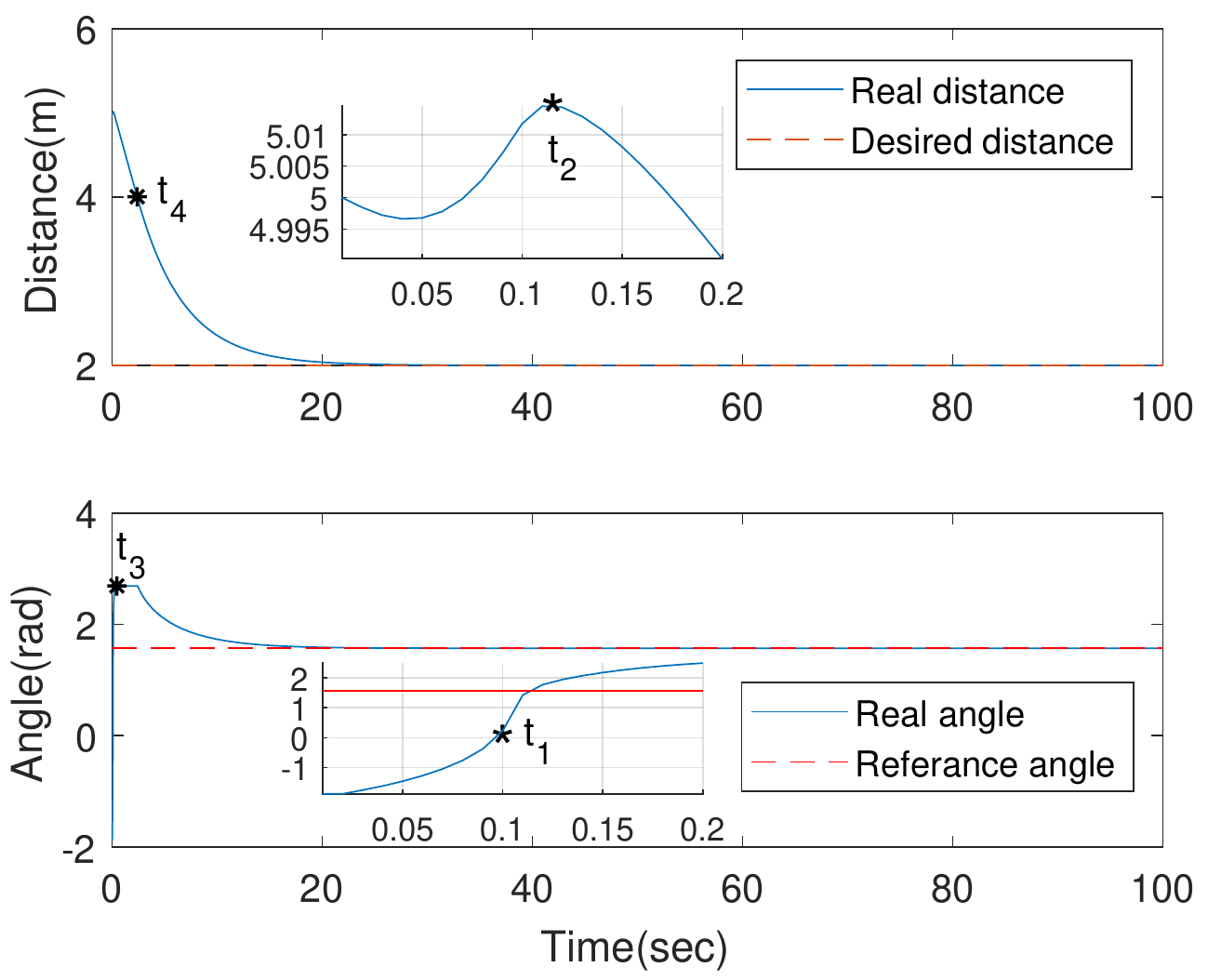}}
	\caption{Distance $d(t)$ and angle $\phi(t)$ versus time.}
	\label{fig4}
\end{figure}

Fig.~\ref{fig4} illustrates the distance $d(t)$ and angle $\phi(t)$ for the initial state $\bm y (t_0)=[7,2,-3\pi/5]'$. The dash lines in Fig.~\ref{fig4} represent the desired radius $r_c=2$ and reference angle $\pi/2$, respectively. It is clear that the target circumnavigation task is eventually completed with zero steady-state error.


 From the partially enlarged view in Fig.~\ref{fig4}, we observe that the angle $\phi(t)$ increases from $\phi(t_0)=-3\pi/5 $ to $\phi(t_1)=0 $ at the initial stage. Then, $\phi(t)$ further increases until $\phi(t_3) = \arccos(-k_2/v_c)$ by crossing $\phi(t_2) = \pi/2$.  From $t_3$ to $t_4$, $\phi(t)$ maintains the value $\arccos(-k_2/v_c)$, and $d(t)$ converges to $r_c$ at the speed $-k_2$. At the time instant $t_4$, it holds that $d(t_4) = r_c + k_3$. Note that, all these observations are consistent with the statements in Subsection \ref{sub2}.

\begin{figure}[t!]
	\centerline{\includegraphics[width=0.8\linewidth]{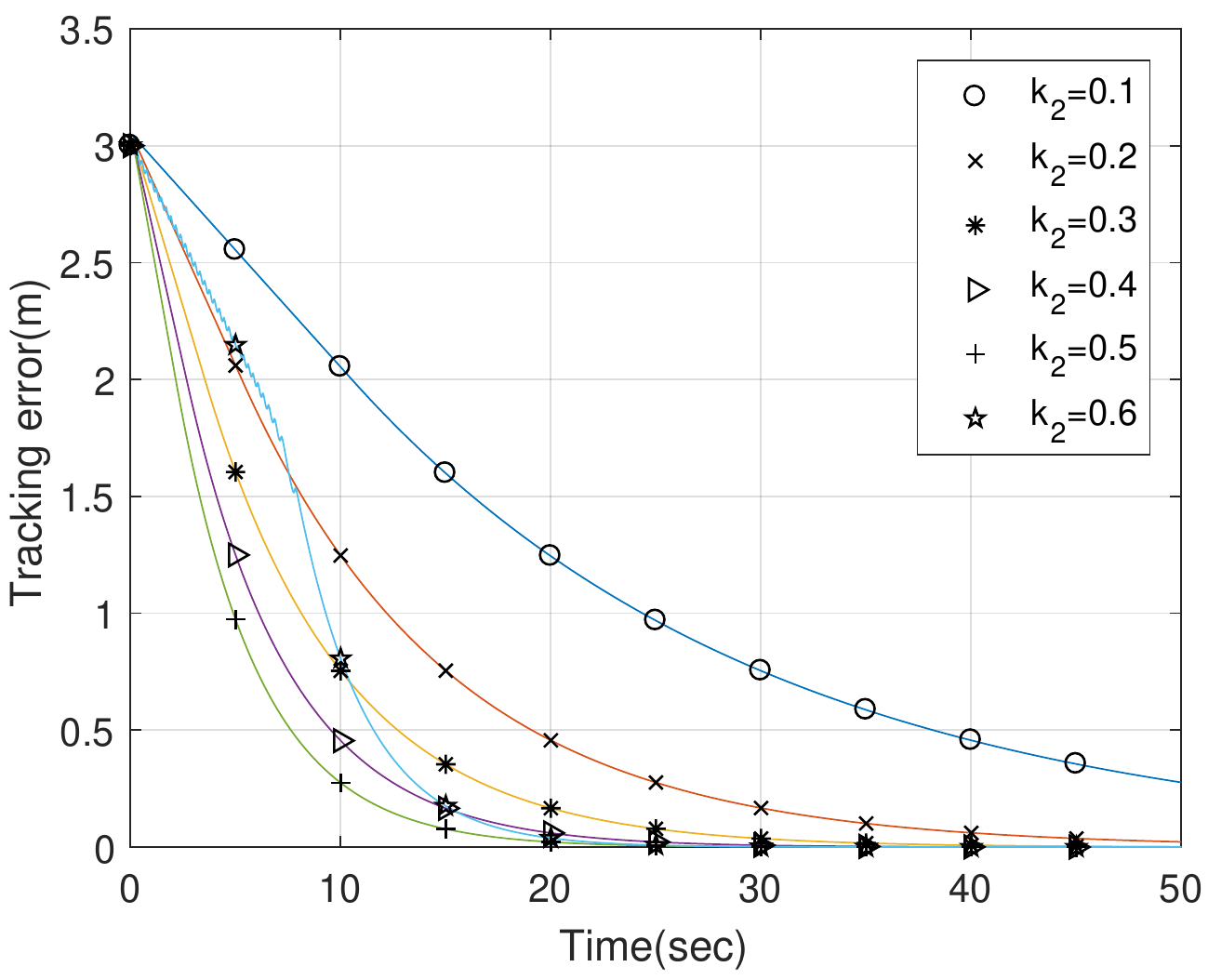}}
	\caption{Distance error $d(t)-r_c$ versus time with different $k_2$.}
	\label{fig6}
\end{figure}


Fig.~\ref{fig6} shows the relationship between the convergence speed and parameter $k_2$. The initial state is $\bm y(t_0) =[7,2,-\pi/2]'$, and the parameters are given in Table \ref{tab2}. If $k_2 > v_c$, e.g., $k_2=0.6$, the oscillation occurs.


\subsection{Target encirclement with smooth patterns} \label{sub13}

In this subsection, we set the reference command as $r(t) = 20+1.8 \sin(0.2t)$. Obviously, $ |\dot r(t)| \le r_v = 0.36$, and $ |\ddot r(t)|\le r_a = 0.072$. $k_2$ is set as $0.1$ and the other parameters are given in Table \ref{tab2}. 
Moreover, the initial state is $\bm y(t_0) = [40,0,\pi/2]'$, satisfying all conditions in Proposition \ref{prop3}. 

\begin{figure}[t!]
	\centerline{\includegraphics[width=0.8\linewidth]{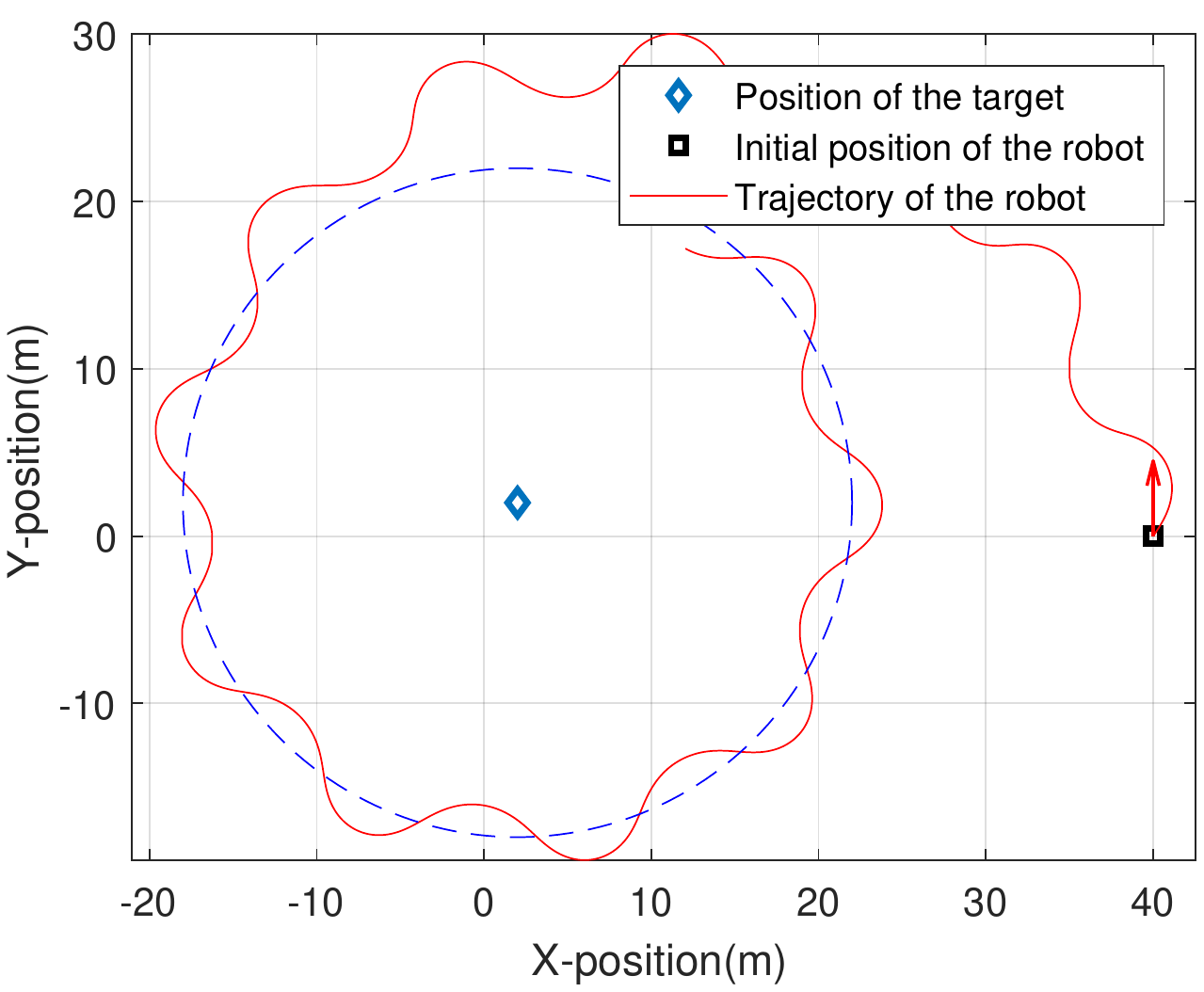}}
	\caption{Trajectory of the robot with  time-varying commands.}
	\label{fig10}
\end{figure}

\begin{figure}[t!]
	\centerline{\includegraphics[width=0.8\linewidth]{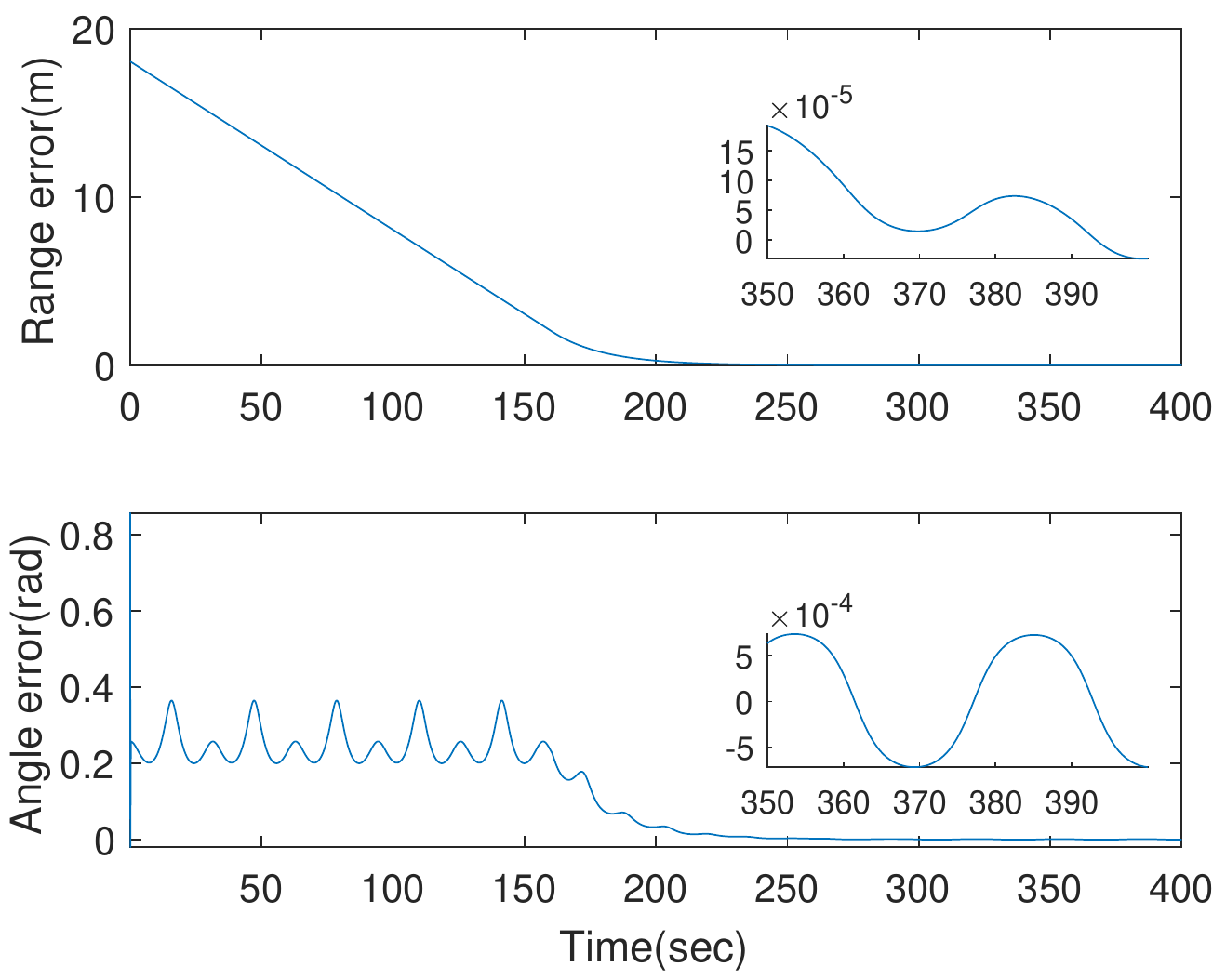}}
	\caption{Distance $d(t)$ and angular $\phi(t)$ versus time with time-varying commands.}
	\label{fig11}
\end{figure}

The results are shown in  Fig.~\ref{fig10} and  Fig.~\ref{fig11}, where the range error is the difference between $d(t)$ and $r(t)$, and the angle error is that between $\phi(t)$ and $\arccos(\dot r(t)/v_c)$. By (\ref{eq5}), $\phi(t) = \arccos(\dot r(t)/v_c)$ implies that $\dot d(t) = \dot r(t)$.  It is clear that $\bm x(t) =[d(t),\dot d(t)]'$ converges to \textit{$\bm x_e(t) =[r(t),\dot r(t)]'$} as  $t\rightarrow \infty$.  

%
%

\subsection{Target encirclement with measurement noises}

%
\begin{figure}[t!]
	\centerline{\includegraphics[width=0.8\linewidth]{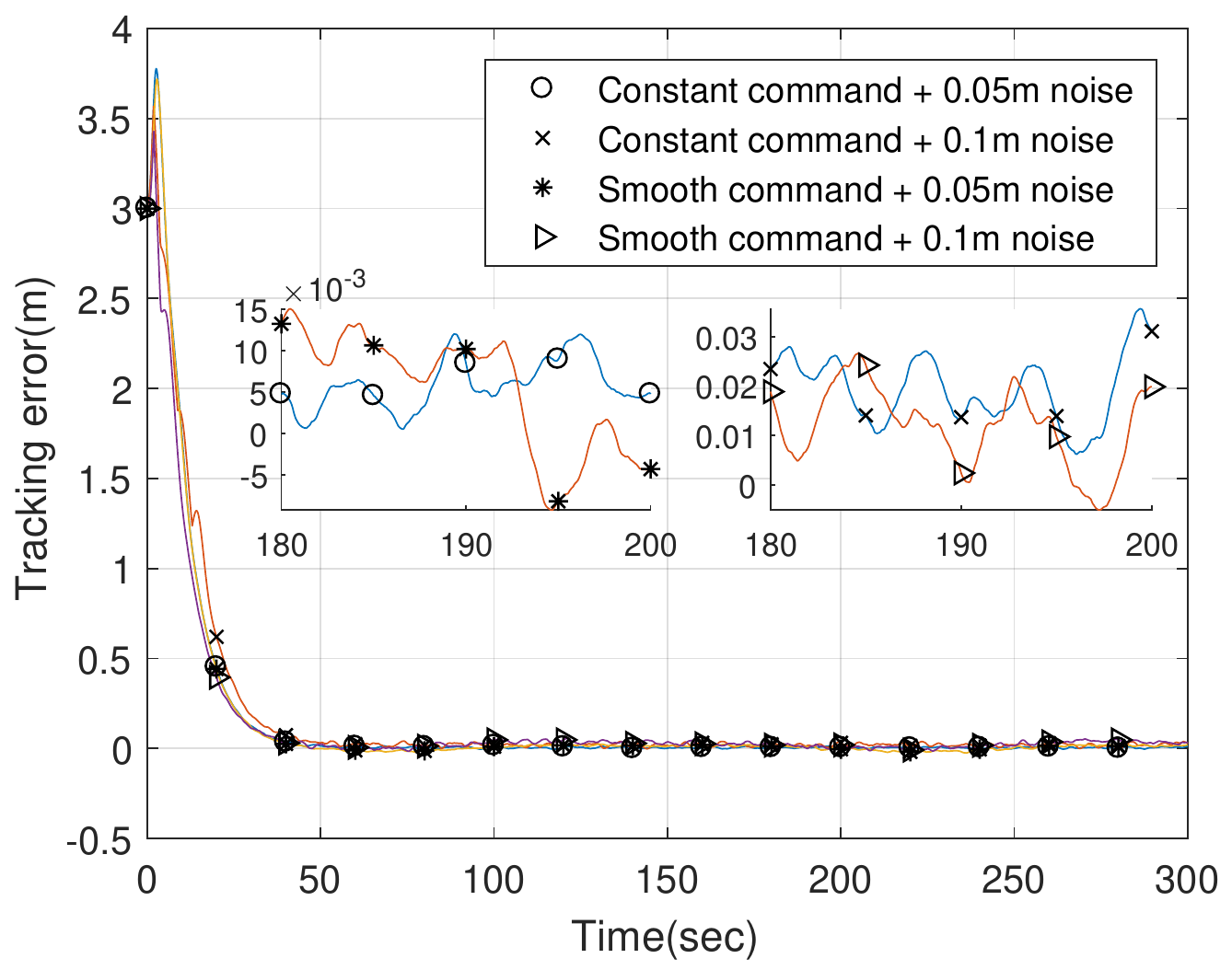}}
	\caption{Tracking errors with different level noises.}
	\label{fig12}
\end{figure}
\begin{table}[t]
	\caption{Parameters of the proposed controller}
	\label{tab3}%
	\centering	
	\begin{tabular}{|c|c|c|c|c|c|c|}	
		\hline
		{Parameter}   &{$k_1$} &{$k_2$}&{$k_3$}&{$h$} &$\varepsilon_1$&$\varepsilon_2$  \\
		\hline
		{Value}         & 1          & 0.25      & 2.0      & 1 & 0.01 & 0.01     \\       
		\hline
	\end{tabular}%
\end{table}%

In this subsection, the range measurement is corrupted by an additive white Gaussian noise, i.e., 
$$d(t) = \Vert \bm p(t) -\bm p_o \Vert _2 + \omega(t),$$ 
where $\omega(t) \sim \mathcal{N}(0,\sigma^2)$. The constant command and time-varying reference command are set as $r_c = 2$ and $r(t)= r_c+0.8*\sin(0.04t)$, respectively.
Fig.~\ref{fig12} shows the results with different noise level: $\sigma = 0.05~\text{and}~ 0.1$, ($2.5\% r_c ~\text{and}~ 5\%r_c$). The initial state is $\bm y(t_0)=[7,2,-\pi/2]$, and the parameters are selected as Table \ref{tab3}. 
From the partially enlarged view of Fig.~\ref{fig12}, we observe that the tracking errors are smaller than the measurement errors.
 This illustrates that the proposed controller is robust against measurement noises. 
 
\subsection{Comparison with the existing methods}

For comparison, we consider the constraint on control output and let $|u(t)| \le \bar u$, where $\bar u = 1$\si{rad/s}  \citep{Matveev2011Range} in this subsection. The comparison methods are the geometrical approach \citep{Cao2015UAV} with parameters $k=1$ and $r_a= 9.95$, the switching approach \citep{Zhang2017Unmanned} with parameter $k = 1.4/r_c$,  and the sliding mode approach \citep{Matveev2011Range} with $\delta=0.83$ and $\gamma = 0.3$. 

\begin{figure}[t!]
	\centerline{\includegraphics[width=0.8\linewidth]{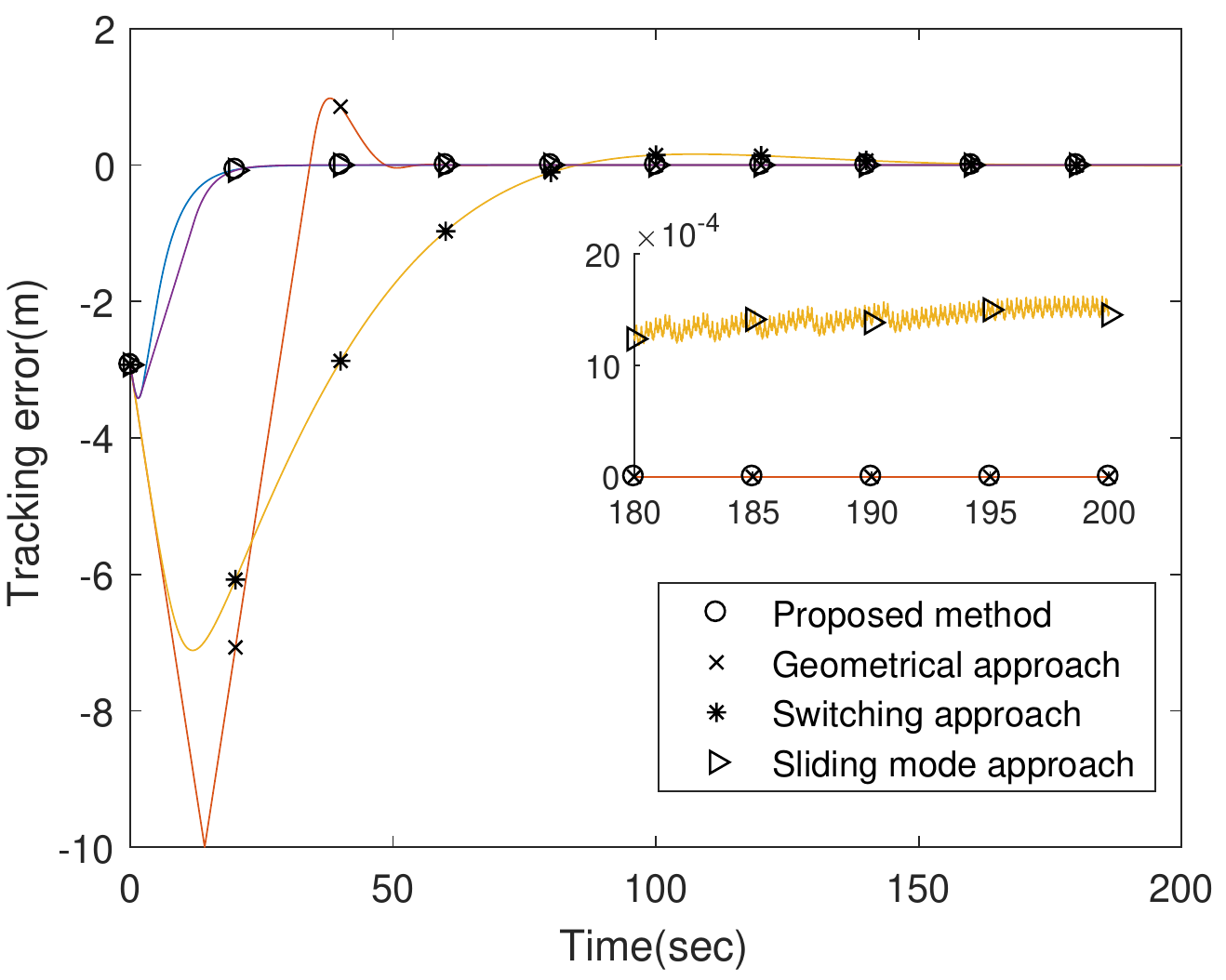}}
	\caption{Comparison of tracking performance with constant reference command. }
	\label{fig14}
\end{figure}

\begin{figure}[t!]
	\centerline{\includegraphics[width=0.8\linewidth]{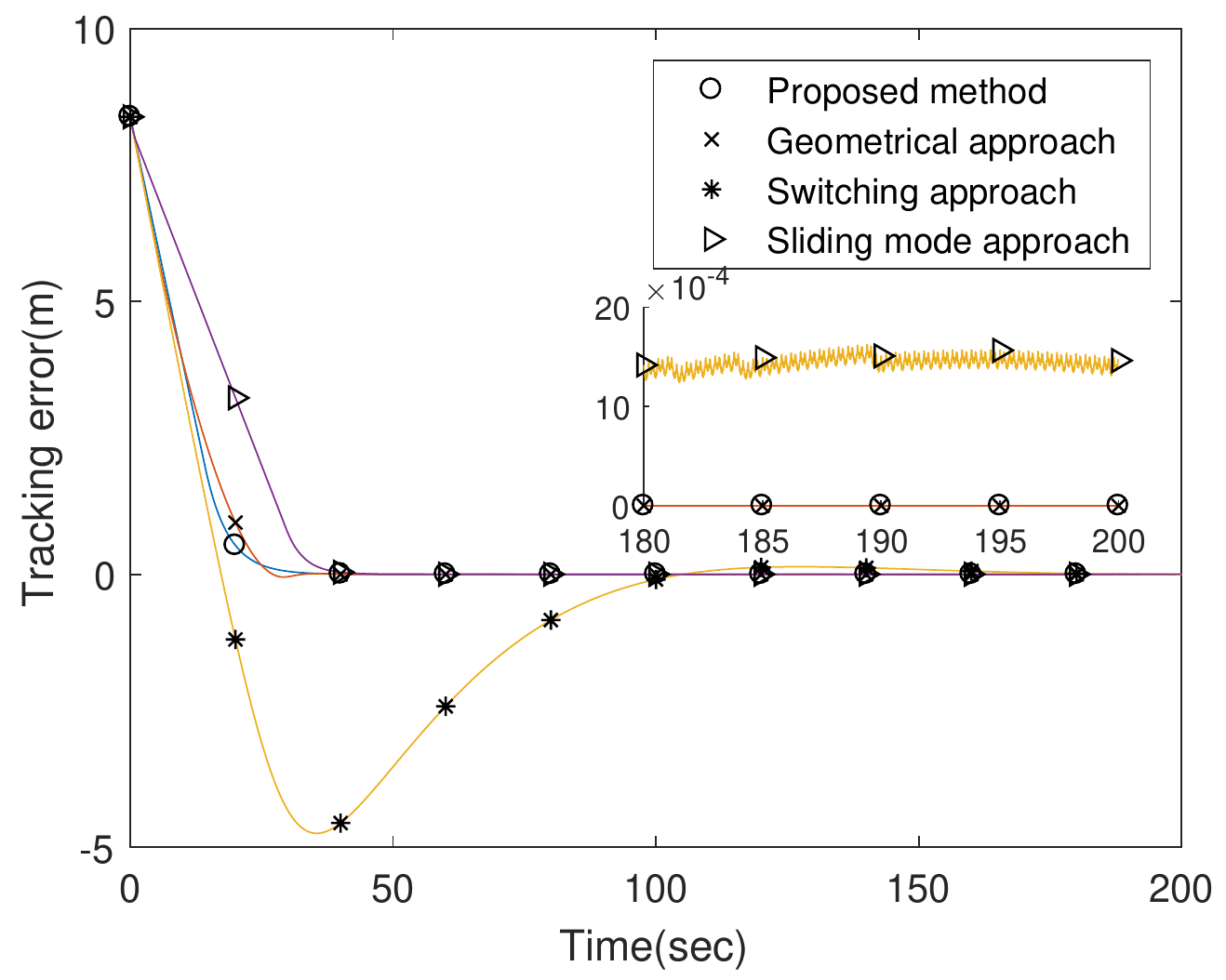}}
	\caption{Comparison of tracking performance with constant reference command.}
	\label{fig17}
\end{figure}

Fig.~\ref{fig14} and Fig.~\ref{fig17} show the results with a constant command $r_c=10$, where the initial states of the robot are $\bm y(t_0)=[7,7,-3\pi/4]$ and $\bm y(t_0)=[15,15,-3\pi/4]$, respectively. The parameters are configured as Table \ref{tab2}.
When $d(t) < r_a$, the control output of the geometrical approach is zero, a large overshoot occurs in this simulation. Furthermore, the convergence speed of the switching approach is slowest.      
Even though the trajectory generated by  the sliding mode approach is similar to that generated by the proposed controller, the sliding mode approach is chattering and has steady-state error, see the partially enlarged view in Fig.~\ref{fig14} and Fig.~\ref{fig17}.  

\begin{figure}[t!]
	\centerline{\includegraphics[width=0.8\linewidth]{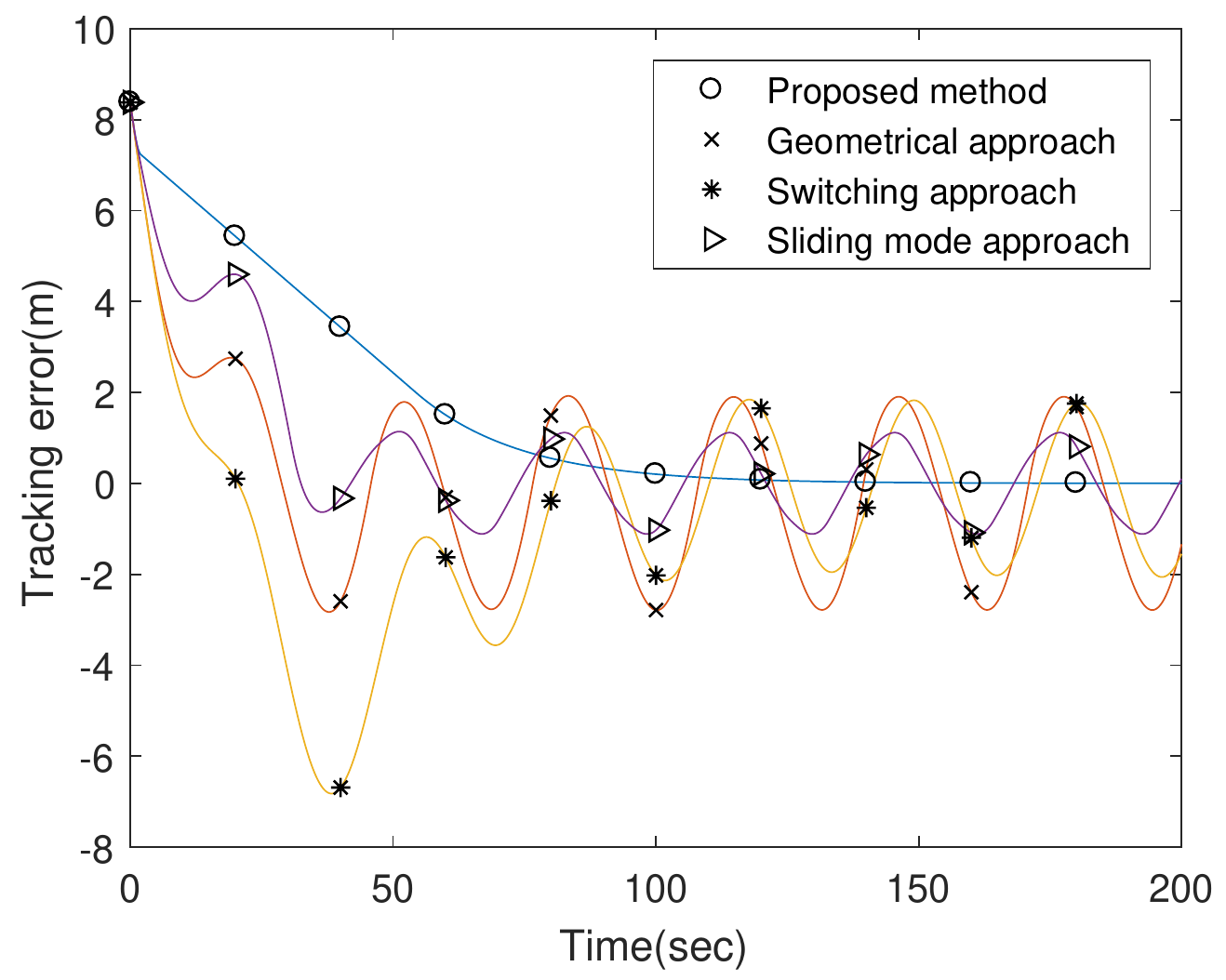}}
	\caption{Comparison of tracking performance with time-varying reference command. }
	\label{fig18}
\end{figure}

Fig.~\ref{fig18} provides the results with smooth patterns. All the methods in \citet{Cao2015UAV,Zhang2017Unmanned,Matveev2011Range} fail to follow the time-varying command $r(t)= 10+1.8 \sin(0.2t)$.

\begin{figure}[t!]
	\centerline{\includegraphics[width=0.8\linewidth]{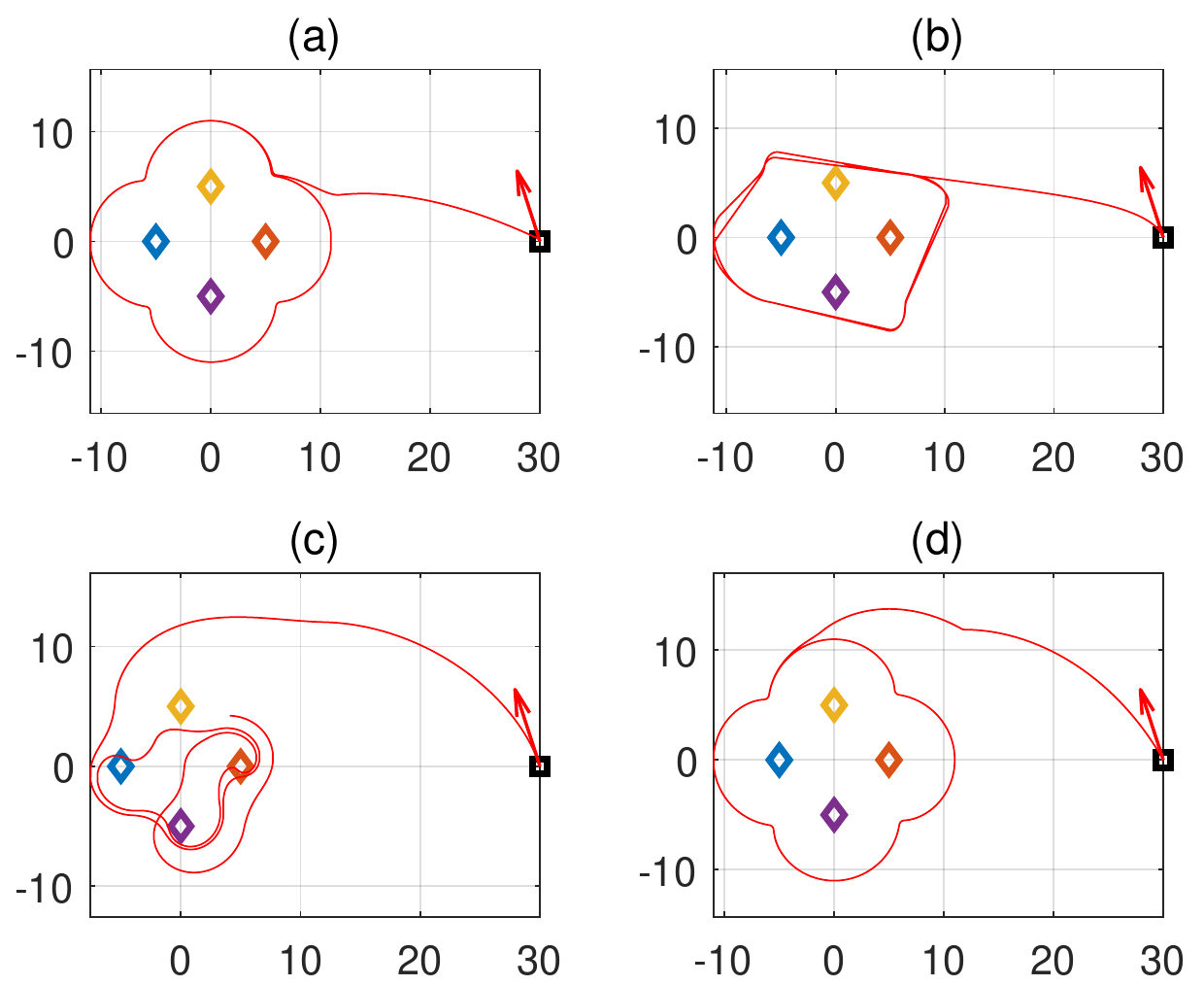}}
	\caption{Multi-target encirclement  under different control methods: (a) proposed method, (b) geometrical approach, (c) switching approach, (d) sliding mode approach.}
	\label{fig19}
\end{figure}

\begin{figure}[t!]
	\centerline{\includegraphics[width=0.8\linewidth]{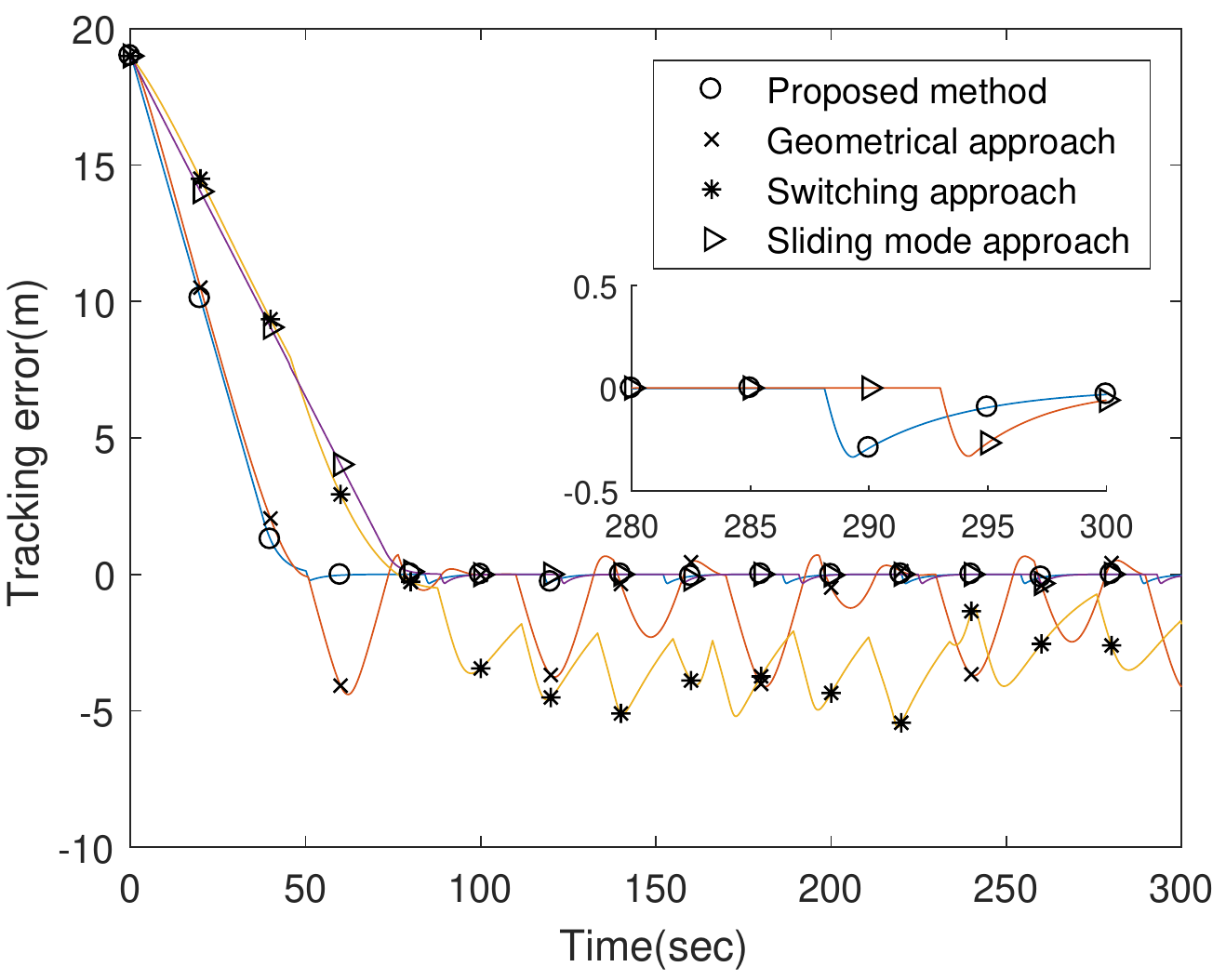}}
	\caption{Comparison of tracking performance. }
	\label{fig20}
\end{figure}

To encircle multiple targets as \citet{MATVEEV201752}, let $$d(t):=\min_i \Vert \bm p(t) - \bm  p_i\Vert_2$$
 where $\bm p_i$ is the position of the $i$-th target. The trajectories and the tracking errors are given in Fig.~\ref{fig19} and Fig.~\ref{fig20}. The proposed method achieves similar performance with that of the sliding mode approach, while either the geometrical approach or the switching approach is unable to complete this task.

Overall, the controller in (\ref{eqc}) can handle several target encircling issues, such as measurement noises, time-varying reference commands, and multiple targets in the simulation. 


\section{Conclusions} \label{sec6}

In this paper, we have proposed a coordinate-free controller to drive a robot to encircle a stationary target with any smooth patterns by only using the range measurements. The proposed controller, which is inspired by the backstepping control method, can guarantee global convergence and exponential stability with zero steady-state error. The simulations validated our theoretical results.

\bibliographystyle{agsm} 
\bibliography{bib/mybib} 

\end{document}